\documentclass{webofc}
\usepackage[varg]{txfonts}   
\newcommand{\nineH}        {$\sqrt{s}~=~0.9$~Te\kern-.1emV\xspace}
\newcommand{\seven}        {$\sqrt{s}~=~7$~Te\kern-.1emV\xspace}
\newcommand{\twoH}         {$\sqrt{s}~=~0.2$~Te\kern-.1emV\xspace}
\newcommand{\twosevensix}  {$\sqrt{s}~=~2.76$~Te\kern-.1emV\xspace}
\newcommand{\five}         {$\sqrt{s}~=~5.02$~Te\kern-.1emV\xspace}
\newcommand{\twosevensixnn}{$\sqrt{s_{\mathrm{NN}}}~=~2.76$~Te\kern-.1emV\xspace}
\newcommand{\fivenn}       {$\sqrt{s_{\mathrm{NN}}}~=~5.02$~Te\kern-.1emV\xspace}

\begin{document}
\title{Hydrodynamization of charm quarks in heavy-ion collisions}
%
%

\author{\firstname{Federica} \lastname{Capellino}\inst{1}\fnsep\thanks{\email{f.capellino@gsi.de}} \and
        \firstname{Andrea} \lastname{Dubla}\inst{2}\fnsep \and
        \firstname{Stefan} \lastname{Floerchinger}\inst{3}\fnsep \and      		\firstname{Eduardo} \lastname{Grossi}\inst{4}\fnsep \and
        \firstname{Andreas} \lastname{Kirchner}\inst{5}\fnsep         \and 		\firstname{Silvia} \lastname{Masciocchi}\inst{1,2}\fnsep        
}
\institute{Physikalisches Institut, Universit{\"a}t Heidelberg, 69120 Heidelberg, Germany
\and
           GSI Helmholtzzentrum f{\"u}r Schwerionenforschung, 64291 Darmstadt, Germany 
\and
           Theoretisch-Physikalisches Institut
Friedrich-Schiller-Universität Jena, 07743 Jena, Germany
\and
			Dipartimento di Fisica, Universit\`a di Firenze and INFN Sezione di Firenze,
50019 Sesto Fiorentino, Italy
\and		Institut f\"ur Theoretische Physik Heidelberg, 69120 Heidelberg, Germany
          }

\abstract{%
  Heavy quarks (i.e. charm and beauty) in heavy-ion collisions are initially produced out of kinetic equilibrium via hard partonic scattering processes. However, recent measurements of anisotropic flow of charmed hadrons pose the question regarding the thermalization of heavy quarks in the quark-gluon plasma (QGP). Exploiting a mapping between transport theory and fluid dynamics, we develop a fluid-dynamic description for charm quarks and show results for charm-hadron momentum distributions. Inspired by recent Lattice-QCD (LQCD) calculations, we show that a late hydrodynamization within the lifetime of the QGP is possible also for beauty quarks. 
}
\maketitle
\section{Introduction}

Charm and beauty quarks in heavy-ion collisions are suitable probes to study the QGP produced in heavy-ion collisions. Due to their large mass, they are produced via hard scatterings occurring at the very beginning of the collision, even before the QGP is created. 
Their dynamics in the QGP is mediated by transport coefficients which encode the microscopic description of the heavy quark-medium interaction. 
In the low transverse momentum region, heavy quarks provide a window to study equilibration processes. 
If particles have enough time to interact with each other, they relax eventually to (local) thermal equilibrium. On the one hand, thermal equilibrium involves (local) chemical equilibrium. This implies that the particle abundance is described by a distribution parametrized by a unique (local) chemical potential $\mu(x)$. Heavy quarks are produced far from chemical equilibrium and, since their number density is much smaller than the one of the light degrees of freedom, they remain out of chemical equilibrium during the full lifetime of the fireball. On the other hand, thermal equilibrium is related to (local) kinetic equilibrium. The latter is achieved if the momentum distribution of the particle is described by a Boltzmann-like
distribution at the same (local) temperature $T(x)$ of the surrounding medium. Although this condition is not fulfilled at the time of the production of the heavy quark-antiquark ($Q\overline{Q}$) pairs, there are indications that it will be by the end of the system evolution. 
Recent measurements of elliptic flow of $\rm D$ mesons and ${\rm J/\psi}$ as a function of $p_{\rm T}$ \cite{ALICE:2020iug} show a positive signal, in line with the one observed for light hadrons. 
This suggests that the hydrodynamization time -- i.e. the typical timescale required for the non-hydrodynamic modes to vanish -- of charm quarks is small enough for them to get dragged along with the QGP. A similar conclusion can be drawn from the most recent LQCD calculations of the heavy-quark spatial diffusion coefficient $D_s$ \cite{Altenkort:2023oms}. 
Motivated by these exciting findings, we address the heavy-quark in-medium dynamics with a fluid-dynamic approach.

\section{Fluid-dynamic equations}

\begin{figure}[hbtp]

\centering
\includegraphics[width=0.48\textwidth]{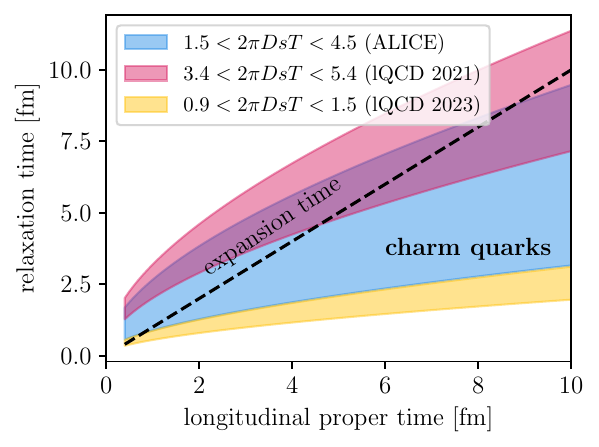}
\includegraphics[width=0.48\textwidth]{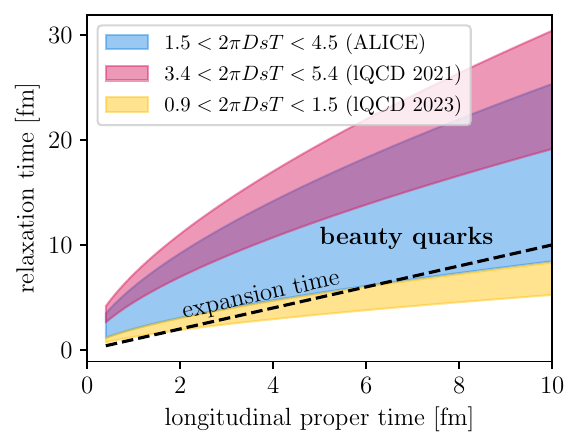}
\caption{Relaxation time of charm quarks (left panel) and beauty quarks (right panel) as a function of proper time in comparison with the typical expansion time of the fluid under the assumption of Bjorken flow. Different colored bands correspond to different values of the spatial diffusion coefficient $D_s$.}
\label{fig:reltimes}
\end{figure}

The fluid-dynamic description of the QGP (see e.g. Ref. \cite{Floerchinger:2018pje}) is here extended to take into account an additional conserved current associated to the heavy-quark number. The conservation of $Q\overline{Q}$ pairs is in fact an effective symmetry of QCD: Due to the heavy quark large mass, the thermal production of a $Q\overline{Q}$ pair is negligible for the temperatures achieved during the fluid-dynamic evolution of the QGP. Furthermore, the annihilation rate of the pairs is negligible within the typical lifetime of the plasma. The associated conserved current is $N^\mu = n u^\mu + \nu^\mu$. It contains a term proportional to the fluid four-velocity $u^\mu$ via the $Q\overline{Q}$-pair density $n$ and a diffusion term $\nu^\mu$ orthogonal to the fluid velocity. Beside the conservation law, 
\begin{equation}
\nabla_\mu N^\mu = 0\,,
\end{equation}
an equation of motion for the diffusion current is needed,
\begin{equation}
\label{eqn:eom_partdiffcurr}
     \tau_n \Delta^\mu_{\,\rho} u^\sigma \partial_\sigma  \nu^\rho + \nu^{\mu}= \kappa_n  \nabla^\mu \left( \frac{\mu}{T} \right)\,,
\end{equation}
where $\Delta^{\mu \nu}=g^{\mu \nu}-u^\mu u^\nu$ is the projector onto the space orthogonal to the fluid velocity and we defined the transverse gradient $\nabla^\mu\equiv\Delta^{\mu\nu}\partial_\nu$. The evolution of the diffusion current is driven by gradients of the $Q\overline{Q}$-chemical potential $\mu$ over temperature $T$. 
The relaxation time $\tau_n$ identifies the hydrodynamization timescale, while $\kappa_n$ regulates the strength of the diffusion process. 
The expressions for the relaxation time $\tau_n$ and the diffusion coefficient $\kappa_n$ were derived in our previous work \cite{Capellino:2022nvf}.
There, an explicit relation between the $\tau_n$ and $D_s$ was found. In Fig. \ref{fig:reltimes}, the comparison between $\tau_n$ and the typical expansion time of the fluid is shown as a function of proper time $\tau$ under the assumption of Bjorken flow. Different colored bands correspond to different input values for $D_s$ given, respectively, by LQCD calculations \cite{Altenkort:2023oms,Altenkort:2020fgs} and fits of multiple transport models to experimental measurements by ALICE \cite{ALICE:2021rxa}. Here we approximate the value of $2\pi D_s T_{\rm c}$ as a constant. The left and right panel show results for charm and beauty quarks, respectively. The relaxation time of charm quarks becomes smaller than the typical expansion time of the fluid very early during the fluid evolution for a broad range of values of $D_s$. This suggests that charm quarks can be meaningfully described by fluid dynamics. The range of applicability of hydrodynamics for beauty quarks seems more limited. However, recent LQCD calculations suggest that a (partial) hydrodynamization of beauty quarks could be possible in the later stages of the fireball.
As for the present work, the focus will be on the study of a hydrodynamic formalism for charm quarks only \cite{Capellino:2023cxe}.   

\section{Charm-hadrons momentum distributions}
\begin{figure}
\centering
\includegraphics[width=0.48\textwidth]{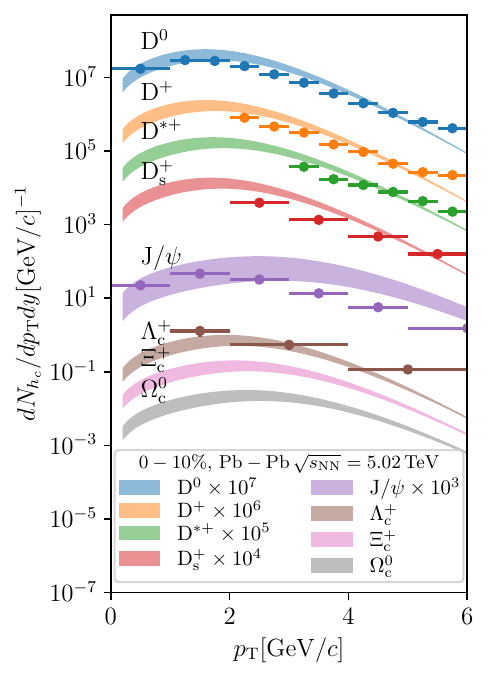}
\includegraphics[width=0.48\textwidth]{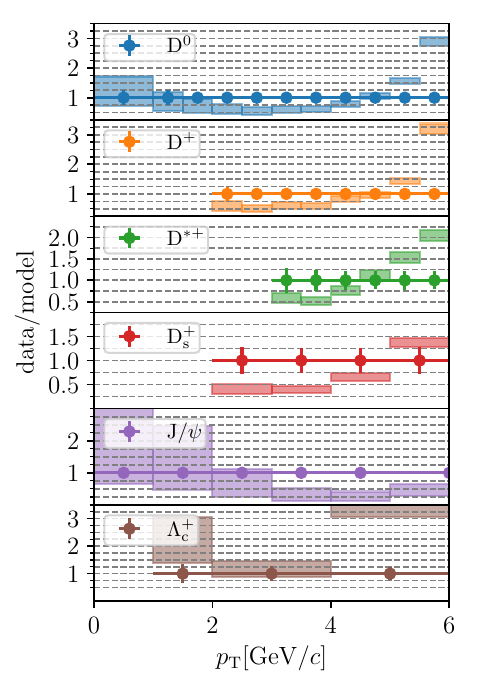}

\caption{Left panel: Results for the momentum distributions of $\mathrm{D^0}$, $\mathrm{D^+}$, $\mathrm{D_s^+}$, $\mathrm{\Lambda_c^+}$ and $\mathrm{J/\psi}$ are shown in comparison with experimental measurements from the ALICE Collaboration \cite{ALICE:2021rxa,ALICE:2023gco,ALICE:2021bib,ALICE:2021kfc}. Right panel: data-to-model ratio plot.
The color bands correspond to a spread on the input value of the spatial diffusion coefficient $D_s$ going from a non-diffusive case ($D_s = 0$) to the upper limit of the LQCD calculations ($2\pi D_s T_c = 1.5$).}
\label{fig:spectra}
\end{figure}

The equations of motion for the stress-energy tensor, charm current and associated dissipative quantities are solved numerically. Since the charm quark number density is small compared to the one of the light degrees of freedom, their contribution to the total energy of the system is neglected. The Equation of State and transport coefficients of the QGP are taken from Ref. \cite{Floerchinger:2018pje}, the charm density is parametrized as in Ref.  \cite{Capellino:2023cxe} and the spatial diffusion coefficient $2\pi D_s T$ is obtained from a linear fit to the LQCD predictions in Ref. \cite{Altenkort:2023oms}.
The initial conditions for the temperature fields are taken using $\mathrm{T_RENTo}$ \cite{Moreland:2014oya} to estimate the initial entropy density deposition in Pb-Pb collisions at 5.02 TeV in the 0-10$\%$ centrality class. The initial distribution of charm quarks scales with the number of binary collisions $n_{\rm coll}$. 
The momentum distributions are obtained employing a Cooper-Frye prescription at a freeze-out temperature of $156.5$ MeV, including resonance decays contributions \cite{Mazeliauskas:2018irt}. The out-of-equilibrium corrections on the freeze-out surface are here not included, and will be object of future developments. In Fig. \ref{fig:spectra} (left panel) our calculations for the spectra of $\mathrm{D^0}$, $\mathrm{D^+}$, $\mathrm{D_s^+}$, $\mathrm{\Lambda_c^+}$ and $\mathrm{J/\psi}$ are shown in comparison with experimental measurements from the ALICE Collaboration \cite{ALICE:2021rxa,ALICE:2023gco,ALICE:2021bib,ALICE:2021kfc}. In the right panel we present a ratio plot with the data to model comparison.
The color bands correspond to a spread of the input value of the spatial diffusion coefficient $D_s$ going from a non-diffusive case ($D_s = 0$) to the upper limit of the LQCD calculations ($2\pi D_s T_{\rm c} = 1.5$). The fluid-dynamic description captures the physics of $\rm D$ mesons up to $p_{\rm T}\sim 4-5$ GeV. The $\mathrm{\Lambda_c^+}$ baryon integrated yield is underestimated, possibly indicating the existence of not-yet-measured resonance states \cite{He:2019tik,He:2019vgs}. The $\mathrm{J/\psi}$ momentum distribution shows a peak for higher $p_{\rm T}$ values with respect to the measured one. This discrepancy might be due to the lack of primordial $\mathrm{J/\psi}$ in our model, which will be studied in the future.
\section{Conclusions and outlook}
This work has shown that a fluid-dynamic description for charm quarks is feasible. Remarkably, the momentum distributions of various charmed hadrons are found to be in agreement with the experimental data in a transverse momentum range up to $4-5$ GeV. 
A consistent way of including the out-of-equilibrium correction at the freeze-out surface has to be developed. To validate the hypothesis of (full) charm thermalization, flow coefficients will be computed and systematically studied against experimental measurements in a continuation of this work.  

\section*{Acknowledgements}

This work is funded via the DFG ISOQUANT Collaborative Research Center (SFB 1225). A.D. is partially supported by the Netherlands Organisation for Scientific Research (NWO) under the grant 19DRDN011, VI.Veni.192.039.


\end{document}